\documentclass[aps,prb,reprint,twocolumn,superscriptaddress,floatfix,citeautoscript]{revtex4-2}
\usepackage[bookmarks=true,colorlinks,linkcolor=OrangeRed,urlcolor=NavyBlue,citecolor=RoyalBlue]{hyperref}
\usepackage{epsfig,amsmath,amssymb,color,comment,physics}
\usepackage[makeroom]{cancel}
\usepackage[caption=false,font=normalsize]{subfig}
\usepackage{mathrsfs}
\usepackage[countmax]{subfloat}
\usepackage[normalem]{ulem}
\usepackage[english]{babel}
\usepackage{dsfont}
\usepackage{placeins}

\usepackage[dvipsnames]{xcolor}
\usepackage{pgfplots}
\pgfplotsset{compat=1.15}
\usetikzlibrary{
        pgfplots.colorbrewer,
    }
\usepackage{multirow} 
\usepackage{graphicx}  

\usepackage{bbold} 
\usepackage{dcolumn}
\usepackage{bm}

\usepackage{siunitx}   
\usepackage{tikz}
\usepackage{calc} 
\usepackage{wasysym} 
\usepackage{mathdesign}

\renewcommand{\vec}[1]{\mathbf{#1}}

\newcommand{\HPF}{\bar{H}_{\text{PF}}}
\newcommand{\HS}{\bar{H}}

\begin{document}

\title{Taming pseudo-fermion functional renormalization for quantum spins: Finite-temperatures and the Popov-Fedotov trick}

\author{Benedikt Schneider}
\affiliation{Department of Physics and Arnold Sommerfeld Center for Theoretical
Physics, Ludwig-Maximilians-Universit\"at M\"unchen, Theresienstr.~37,
80333 Munich, Germany}
\affiliation{Munich Center for Quantum Science and Technology (MCQST), 80799 Munich, Germany}
\author{Dominik Kiese}
\affiliation{Institute for Theoretical Physics, University of Cologne, 50937 Cologne, Germany}
\affiliation{Center for Computational Quantum Physics, Flatiron Institute, 162 5th Avenue, New York, NY 10010, USA}
\author{Björn Sbierski}
\affiliation{Department of Physics and Arnold Sommerfeld Center for Theoretical
Physics, Ludwig-Maximilians-Universit\"at M\"unchen, Theresienstr.~37,
80333 Munich, Germany}
\affiliation{Munich Center for Quantum Science and Technology (MCQST), 80799 Munich, Germany}

\begin{abstract}
The pseudo-fermion representation for $S=1/2$ quantum spins introduces unphysical states in the Hilbert space which can be projected out using the Popov-Fedotov trick. However, state-of-the-art implementation of the functional renormalization group method for pseudo-fermions have so far omitted the Popov-Fedotov projection. Instead, restrictions to zero temperature were made and absence of unphysical contributions to the ground-state was assumed. 
We question this belief by exact diagonalization of several small-system counterexamples where unphysical states do contribute to the ground state.
We then introduce Popov-Fedotov projection to pseudo-fermion functional renormalization, enabling finite temperature computations with only minor technical modifications to the method. At large and intermediate temperatures, our results are perturbatively controlled and we confirm their accuracy in benchmark calculations. At lower temperatures, the accuracy degrades due to truncation errors in the hierarchy of flow equations. Interestingly, these problems cannot be alleviated by switching to the parquet approximation. We introduce the spin projection as a method-intrinsic quality check. We also show that finite temperature magnetic ordering transitions can be studied via finite-size scaling.
\end{abstract}

\date{\today}
\maketitle

\section{Introduction}
\label{sec:introduction}
Many quantum systems of current interest, ranging from frustrated magnets \cite{auerbach_interacting_1994} to Rydberg atom arrays \cite{browaeys_many-body_2020} can be described by Hamiltonians consisting of spin $S=1/2$ operators $\mathbf{S}=(S^x,S^y,S^z)$ fulfilling the standard $\mathfrak{su}(2)$ spin algebra \cite{auerbach_interacting_1994}. In theoretical treatments, it is often useful to switch to an auxiliary particle representation of the spin operator. An established representation in terms of spinful fermions annihilated by $f_{\alpha=\uparrow,\downarrow}$ goes back to Abrikosov \cite{abrikosov_electron_1965},
\begin{align}
    S^\mu \xrightarrow{} \Bar{S}^\mu = \frac{1}{2} \sum_{\alpha,\alpha'} f^\dagger_\alpha \sigma^\mu_{\alpha\alpha'} f_{\alpha'}. \label{eq:pf}
\end{align}
Here, $\sigma^\mu$ are Pauli matrices $(\mu=x,y,z)$ and an overbar indicates an operator in the fermionic Hilbert space \footnote{We skip the overbar for fermionic operators like $f_\alpha$ or $n_\alpha=f_\alpha^\dagger f_\alpha$ where no confusion is possible.}.

The pseudo-fermion (pf) representation \eqref{eq:pf} allows for a variety of applications. For example, it is one of the pillars of the theory of spin-fractionalization and spin-liquids \cite{wen_quantum_2007} where pf mean-field states are used to describe highly entangled paramagnetic ground states of frustrated spin systems. On the other hand, the pf representation has been used extensively for numerical methods as it allows to transfer the well-developed diagrammatic toolbox for interacting fermions in equilibrium \cite{negele_quantum_2018} to quantum spins. These tools are based on the Wick theorem and perturbation theory. Two popular examples for more advanced methods are the diagrammatic Monte Carlo \cite{kulagin_bold_2013,kulagin_bold_2013-1,huang_spin-ice_2016}(pf-diagMC) and the functional renormalization group (pf-fRG) \cite{reuther_j_2010} in vertex expansion. Whereas the first method samples diagrams of a perturbative series in $J/T$ ($J$ is the exchange coupling and $T$ the temperature) \cite{van_houcke_diagrammatic_2010}, the second rests on a hierarchy of flow equations for pf vertex functions which flow under the variation of the regularized bare propagator \cite{kopietz_introduction_2010,metzner_functional_2012}. Despite the necessary truncation of this hierarchy, resulting end-of-flow correlation functions contain infinite-order re-summations of certain diagram classes. 

Crucially, there is a well-known problem that appears whenever the pf representation \eqref{eq:pf} is used: While the left hand side acts in the two-dimensional spin Hilbert space spanned by $\{\ket{\uparrow},\ket{\downarrow}\}$, the basis of the right hand side's Hilbert space is extended, $\{\ket{\uparrow},\ket{\downarrow},\ket{\uparrow\downarrow},\ket{0}\}$ and the pf spin operator is $\Bar{S}^\mu = \mathrm{diag}(S^{\mu},0,0)$. It faithfully represents a spin $S=1/2$ operator only on the physical subspace $\{\ket{\uparrow},\ket{\downarrow}\}$ while it acts as a $S=0$ operator on the empty and doubly-occupied subspaces. As the mentioned fermionic methods are applied in thermal equilibrium, any occupation of the unphysical $S=0$ subspaces will lead to differences between the physical $S^\mu$ correlation functions and the pf ones (using $S^\mu \xrightarrow{} \Bar{S}^\mu$) as well as between the associated partition functions $Z$ and $\Bar{Z}$. 

Fortunately, this problem can be circumvented with a trick found by Popov and Fedotov (PF) in the late 1980s \cite{popov_functional-integration_1988,prokofev_popov-fedotov_2011-1}. We will review the details of the PF trick in Sec.~\ref{sec:Popov}. In its most simple incarnation it amounts to the addition of an imaginary valued chemical potential term to the pf Hamiltonian, $\Bar{H} \rightarrow \Bar{H}+\bar{H}_\text{PF}$, where
\begin{align}
     \Bar{H}_{\text{PF}} = \frac{i\pi T}{2}(n_\downarrow + n_\uparrow  -1) \label{eq:PopovPotential}
\end{align}
and $n_\alpha=f^\dagger_\alpha f_\alpha$ is the pf number operator.

Whereas the PF trick is routinely and straightforwardly employed in pf-diagMC calculations \cite{kulagin_bold_2013,kulagin_bold_2013-1,huang_spin-ice_2016}, this is not the case in the pf-fRG literature where only very limited attention has been paid to the subject \cite{reuther_j_2010,reuther_frustrated_2011,roscher_cluster_2019}. The justifying narrative for this omission in state-of-the-art pf-fRG \cite{reuther_functional_2011,reuther_finite-temperature_2011,singh_relevance_2012,reuther_cluster_2014,iqbal_functional_2016,baez_numerical_2017,iqbal_quantum_2019,thoenniss_multiloop_nodate, kiese_multiloop_2022}
is that unphysical $S=0$ states would only occur at energies above the ground state energy and consequently PF projection would be unnecessary if calculations are restricted to $T=0$. However, this state of affairs comes with a number of problems:

(i) In the following Sec.~\ref{sec:Constraint}, we show that even simple and generic frustrated spin systems like the antiferromagnetic (AFM) Heisenberg trimer have a pf ground state with sizable occupation of unphysical $S=0$ states if PF projection is omitted. Even if these clusters are usually not the focus of pf-fRG applications, they are basic building blocks of highly relevant lattices like triangular or Kagome. For AFM Heisenberg Hamiltonians on these lattices it is thus questionable if ground states in pf representation without PF projection are indeed in the physical subspace. As these systems have been studied amply with pf-fRG \cite{reuther_functional_2011,suttner_renormalization_2014,buessen_competing_2016,iqbal_spin_2016,thoenniss_multiloop_nodate,kiese_pinch-points_nodate}, this also questions the quantitative accuracy of these state-of-the-art pf-fRG results.

(ii) Even for systems where the pf ground state is faithful, the unavoidable truncation of the hierarchy of fRG flow equations is an uncontrolled approximation at $T=0$, where, in the absence of magnetic fields $h$, neither $J/T$ nor $J/h$ can be used to perturbatively justify the truncation. Existing arguments appealing to the correctness of commonly employed pf-fRG truncations in the large $S$ or large $N$ limit \cite{baez_numerical_2017,buessen_functional_2018} [in the sense of generalizing $SU(2)$ to $SU(N)$] are not rigorously helpful in the case $S=1/2$, $N=2$ most relevant in applications. 
At best, it is the presence of the Matsubara cutoff scale $\Lambda$ which perturbatively controls the pf-fRG: As $\Lambda$ flows from infinity to zero, only ordering tendencies extracted at $\Lambda \gtrsim J$ tend to be reliable, a point of view not emphasized in the literature before \footnote{Details will be covered in a forthcoming publication}. However, as the physical model is only recovered as $\Lambda \rightarrow 0$, quantitatively reliable results for observables cannot be guaranteed by this argument. 
Another practical difficulty of the pf-fRG at $T=0$ is the continuous nature of Matsubara frequencies, requiring major numerical efforts for a stable and reliable solution of the flow equations \cite{ritter_benchmark_2022}, in particular if multiloop schemes are involved \cite{kiese_multiloop_2022,thoenniss_multiloop_nodate}. 

In this work, we tame problem (i) and (ii) of the pf-fRG by implementing the PF trick in the pf-fRG framework. We term the resulting method ppf-fRG. We show that this is possible with only minor technical modifications (see Sec.~\ref{sec:correlators}) so that most methodological achievements from the last decade  \cite{baez_numerical_2017,hering_functional_2017,buessen_functional_2019,classen_competing_2019,thoenniss_multiloop_nodate,kiese_multiloop_2022,gresista_moments_2022} can be seamlessly adapted.
As a result, the ppf-fRG leverages the pf-fRG to finite temperatures $T$ where quantum and thermal fluctuations compete, and the numerical implementation simplifies by the discrete nature of Matsubara frequencies. In addition, we show that the ppf-fRG is now perturbatively justified for small $J/T$ and equipped with a method-intrinsic gauge for the quality of the results. 

In Sec.~\ref{sec:Results} we benchmark the ppf-fRG using small spin clusters. Although exact results are obtained trivially for these systems, diagrammatic approaches are invoked already at their full complexity, making spin clusters a valuable testbed. We show that the ppf-fRG indeed yields quantitatively reliable results at large and moderate $T$ but cannot be trusted for $T$ smaller than about a third of $J$, at least for the models studied here. Surprisingly, we find that this situation cannot be improved by considering solutions of the parquet approximation. The latter is known to be in equivalence with the loop-converged limit of the multiloop fRG \cite{kugler_multiloop_2018,kugler_multiloop_2018-1,kugler_derivation_2018} in which the two-particle vertex is correctly obtained up to errors of order $J^{4}$, i.e., one order higher than in one-loop fRG.

By applying the ppf-fRG to a translation invariant three-dimensional Heisenberg magnet on the cubic lattice, we show that the study of finite-temperature transitions into symmetry-broken (magnetic) phases becomes possible. In Sec.~\ref{sec:conclusion}, we conclude and also relate our results to the recently developed pseudo-Majorana fRG (pm-fRG) \cite{niggemann_frustrated_2021} where \textit{by construction} no unphysical sectors exist in the fermionic Hilbert space.

\section{Spin constraint violation at $T \geq 0$}
\label{sec:Constraint}

\begin{figure}
    \centering
    \subfloat{%
\resizebox{\columnwidth}{!}{%
\hspace{5mm}
\includegraphics[width = 86mm]{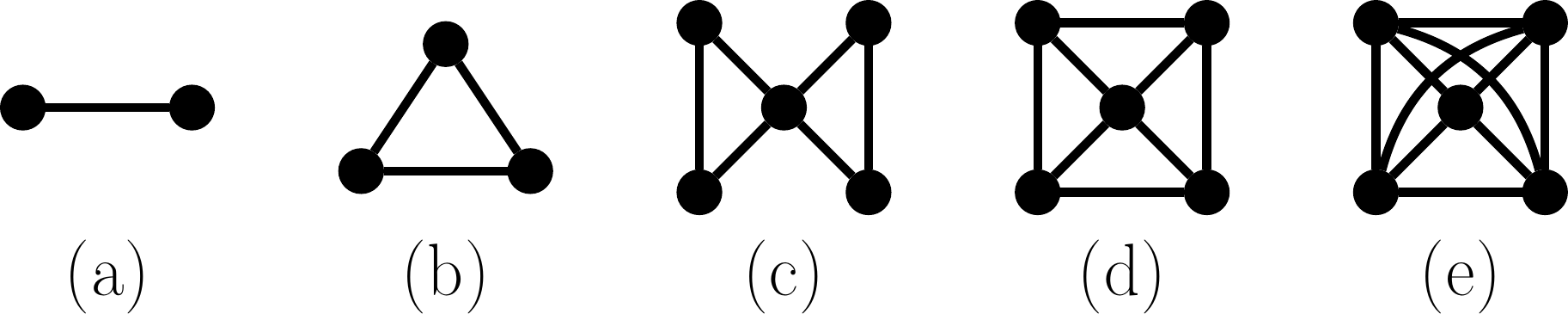}
}
}    
    
\subfloat{%
    \includegraphics[width = 86mm]{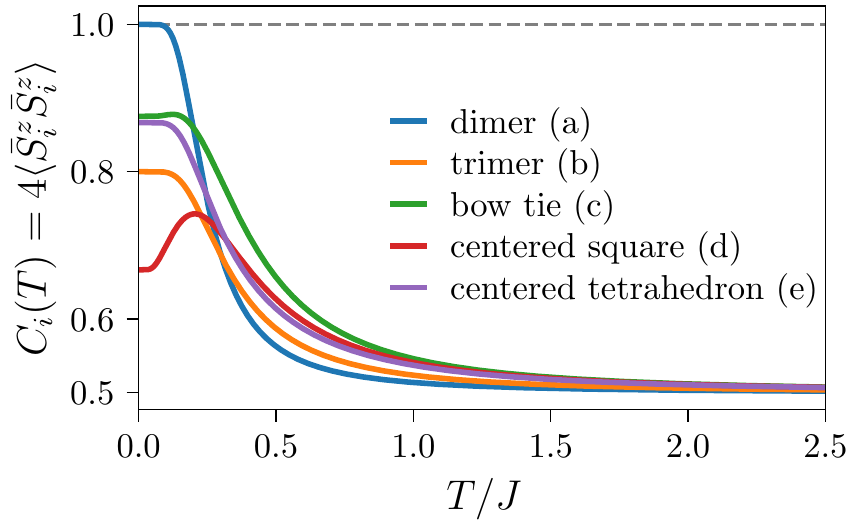}}\\

    \caption{Spin projection $C_i(T)=4\langle \bar{S}^z_i \bar{S}^z_i \rangle$ over temperature calculated via exact diagonalization of the pf Hamiltonian $\bar{H}$ for Heisenberg spin-clusters of $N=2,3,5$ sites as shown in the legend. A bond represents an anti-ferromagnetic Heisenberg coupling $J=1$. For the bow tie and the centered square, with inequivalent sites, the site-label $i$ refers to the center spin. For all clusters except the dimer, the pf ground state manifold contains unphysical states and the $T\rightarrow0$ limit of $C_i(T)$ reduces from the physical value of unity (dashed line) to $\frac{4}{5}$ for the trimer, $\frac{7}{8}$ for the bow tie, $\frac{2}{3}$ for the centered square and $\frac{13}{15}$ for the centered tetrahedron, respectively.
    }
    \label{fig:ClusterConstraints}
\end{figure}

In this section we substantiate our claim that there exist simple AFM spin clusters with a ground state of the pf Hamiltonian $\Bar{H}$ that -- without PF projection -- partially resides outside of the physical $S=1/2$ sector. We use exact diagonalization (ED) of the pf Hamiltonian $\Bar{H}$ corresponding to the spin Hamiltonian $H$ according to Eq.~\eqref{eq:pf}. Our examples serve to demonstrate the need for PF projection along with the pf representation [cf.~(i) in Sec.~\ref{sec:introduction}], but the small system sizes allow to disregard problems related to the solution of $\Bar{H}$, i.e.~the truncation of the pf-fRG [cf.~(ii)]. 

The most elementary example is the AFM Heisenberg trimer with $N=3$ spins coupled all-to-all ($J=1$), see (b) in Fig.~\ref{fig:ClusterConstraints},
\begin{align}
    H_\mathrm{trimer} = J(\vec{S}_1\cdot \vec{S}_2+\vec{S}_2\cdot \vec{S}_3+ \vec{S}_3\cdot \vec{S}_1).\label{eq:trimer}
\end{align}
Its degenerate physical ground state manifold is described by two of the spins forming a singlet and the third spin being in any other state. By using the pf representation \eqref{eq:pf}, $H_\mathrm{trimer} \rightarrow \Bar{H}_\mathrm{trimer}$, the state with the third spin in an unphysical $S=0$ configuration has the same energy ($E= -3J/4$) as the physical ground state and is therefore equally populated at $T=0$. To quantify this further we consider
\begin{equation}
    C_i(T) = 4\langle \bar{S}^z_i \bar{S}^z_i \rangle = \langle \Bar{P_i} \rangle,
\end{equation} 
first studied in Ref.~\cite{thoenniss_multiloop_nodate} in the context of pf-fRG. On the right side, the projector to the local $S=1/2$ sector is
\begin{equation}
\Bar{P_i}=n_{i\uparrow}+n_{i\downarrow} - 2n_{i\uparrow}n_{i\downarrow}. \label{eq:P}
\end{equation}
Only if the (local) pf configuration is entirely in the physical subspace, $\bar{S}^z_i$ would be a faithful $S=1/2$ operator squaring to $1/4$ and $C_i(T)$ would be unity. Thus $C_i(T)$ can be interpreted as a measure for the suppression of unphysical states and we refer to it as {\it spin projection}. Returning to the trimer, we obtain $C_i(T=0)=4/5<1$ signaling the presence of an admixture of unphysical $S=0$ states to the ground state manifold. We plot $C_i(T)$ for $\Bar{H}_\mathrm{trimer}$ over a range of temperatures, see Fig.~\ref{fig:ClusterConstraints}. The spin projection decreases with increasing $T$ reaching $C_i(T\rightarrow \infty) = 1/2$, indicating that in the $T=\infty$ state on the pf Hilbert space the occupation of unphysical states equals the occupation of physical states. 

Besides the trimer we found numerous other AFM Heisenberg spin clusters with $N=5$ where unphysical states poison the pf ground state. This behavior can be generally observed when a physical system cannot lower its ground state energy by adding one more spin and is thus paradigmatic for frustrated systems. In that case, preparing the additional site in a physical $S = 1/2$ or an unphysical $S=0$ state gives the same ground state energy. In Fig.~\ref{fig:ClusterConstraints}, we show results for the bow tie (c), centered square (d) and centered tetrahedron (e), which are qualitatively similar to the trimer case (b). Since these shapes (including the trimer) are basic building blocks of the triangular, Kagome, face-centered cubic and  centered pyrochlore lattice \cite{nutakki_frustration_nodate}, respectively, it is questionable if ground states of AFM Heisenberg pf Hamiltonians $\Bar{H}$ on these lattices reside entirely in the physical subspace. Finally, from the clusters considered in Fig.~\ref{fig:ClusterConstraints}, only the $N=2$ dimer (a) has a pf ground state in the physical subspace.

\section{Popov-Fedotov trick: A review}
\label{sec:Popov}

As discussed in Sec.~\ref{sec:introduction}, the spin operator $\Bar{S}^\mu$ in pf representation \eqref{eq:pf} acts like a spin-$1/2$ operator on the subspace $\{\ket{\uparrow},\ket{\downarrow}\}$ but like a spin-$0$ operator on the subspaces spanned by $\ket{\uparrow\downarrow}$ and $\ket{0}$, respectively. Thermal occupation of the latter two sectors of the fermionic Hilbert space will thus compromise the equivalence between the spin and pf partition functions, $Z\neq \Bar{Z} \equiv \mathrm{tr}\,e^{-\beta \bar{H}}$, respectively.
A straightforward projection to the physical subspace using $\Bar{P}=\Pi_{i=1,...,N}\Bar{P_i}$ with local projectors $\Bar{P}_i$ from Eq.~\eqref{eq:P} becomes unpractical for a large number of spins $N$. 
 
A more feasible projection scheme was originally proposed by Popov and Fedotov \cite{popov_functional-integration_1988} and later generalized by Prokof'ev and Svistunov  \cite{prokofev_popov-fedotov_2011-1}. It amounts to replacing $\Bar{H} \rightarrow \Bar{H}+\bar{H}_\text{PF}^{\phi_{1,2}}$ where
 \begin{align}
     \Bar{H}_{\text{PF}}^{\phi_{1,2}} = i T \left[ n_\downarrow n_\uparrow  \phi_1  + (n_\uparrow-1)(n_\downarrow-1)\phi_2 \right]
     \label{eq:PF_general}
 \end{align}
with $\phi_{1,2}\in\mathbb{R}$ and the constraint $e^{i\phi_1} + e^{i\phi_2} = 0$. Now, the partition function in the pf Hilbert space can be split into a sum over the purely physical part of the Hilbert space (equivalent to $Z$ since $\Bar{H}_\text{PF}^{\phi_{1,2}}=0$ in the $S=1/2$ sector) and a part containing the expectation values of a product state with $\mathcal{N}=1,2,...,N$ unphysical contributions,
\begin{align}
   \Bar{Z}=\mathrm{tr}(e^{-\beta \Bar{H}-\beta\bar{H}_\text{PF}^{\phi_{1,2}} }) &=  Z +  \sum_{\mathcal{N}=1}^N\sum_{\xi_{\mathcal{N}}}(Z_{\xi_\mathcal{N}} \prod_{j \in \xi_\mathcal{N}}F_j^{\phi_{1,2}}).
   \label{eq:PF trick}
\end{align}
Here, $\xi_\mathcal{N}$ counts different configurations of sites with unphysical spin and $Z_{\xi_\mathcal{N}}$ is the physical partition function of the subsystem with the $\mathcal{N}$ unphysical sites removed. Since $\Bar{H}$ acts trivially on unphysical states, we can compute the local trace $F_j^{\phi_{1,2}}$ over the unphysical states at site $j$ explicitly,
\begin{align}
    F_j^{\phi_{1,2}} &= \bra{0}_j e^{-\beta \Bar{H}_\text{PF}^{\phi_{1,2}} } \ket{0}_j + \bra{\uparrow\downarrow}_j e^{-\beta \Bar{H}_\text{PF}^{\phi_{1,2}} }\ket{\uparrow\downarrow}_j\\
    &= e^{i\phi_1} + e^{i\phi_2} = 0
\end{align}
and Eq.~\eqref{eq:PF trick} yields $\Bar{Z}=Z$. A similar analytic argument shows the faithfulness of spin correlation functions computed using the pf representation with the PF term \eqref{eq:PF_general}, in particular $C_i(T)=1$ for all temperatures.

In the rest of the paper, we make the choice $\phi_1=-\phi_2=  \frac{\pi}{2}$. This reduces Eq.~\eqref{eq:PF_general} to Eq.~\eqref{eq:PopovPotential}, taking the simple non-interacting form of a potential which is, however, imaginary.

We emphasize that the PF trick applies also in the limit $T \rightarrow 0$ as easily seen from the fact that in the partition function (for which the PF term is rigorously defined) the Hamiltonian is multiplied by $\beta=1/T$ and the PF contribution amounts to phase factors independent of $T$. 

\section{Symmetries, correlation functions and the fRG}
\label{sec:correlators}

In this technical section we review the symmetries of the generic pf Hamiltonian $\Bar{H}$ following Buessen et al.~\cite{buessen_functional_2019}. We then discuss the necessary changes enforced by the addition of the PF potential term \eqref{eq:PopovPotential}. The resulting modifications in the parametrization of correlation- (and vertex-) functions are minor and can easily be implemented into established numerical codes solving the pf-fRG flow equations \cite{kiese_multiloop_2022,buessen_spinparser_2022}. We call the resulting pf-fRG formalism including the PF potential ppf-fRG and restrict our focus to one-loop evaluation of the flow equations (including Katanin truncation) and an iterative solution of the parquet approximation, ppf-PA. 

\subsection{Symmetries of the pf Hamiltonian}
In analogy to the pf-fRG literature, we restrict our discussion to spin Hamiltonians with two-spin interaction across bonds $(i,j)$ and disregard magnetic fields,
\begin{align}
    H = \sum_{(i,j)}\sum_{\mu,\nu=x,y,z} J_{(i,j)}^{\mu\nu} S^{\mu}_i S^{\nu}_j.
\end{align}
Using Eq.~\eqref{eq:pf}, $S_i^\mu\xrightarrow{}\bar{S}^\mu_i$, and the PF potential term $\HPF$ from Eq.~\eqref{eq:PopovPotential}, we consider the pf Hamiltonian
\begin{equation}
    \HS + \HPF = \sum_{(i,j)}\sum_{\mu,\nu} J_{(i,j)}^{\mu\nu} \bar{S}^{\mu}_i \bar{S}^{\nu}_j +\frac{i\pi}{2\beta}\sum_{j}(n_{\uparrow j} + n_{\downarrow j}  -1).
\end{equation}
For $\HS$ alone, Buessen et al.~\cite{buessen_functional_2019} have discussed the following properties: $\HS$ is hermitian (H) and symmetric with respect to  local $\text{U}(1)$, local particle-hole (lPH) and (anti-unitary) time-reversal (TR) transformation. These symmetries act on $\mathcal{F}_{i\alpha} \equiv (f_{i \alpha}^{\dagger},f_{i \alpha} )^T$ as follows,
\begin{align}
\mathcal{F}_{i\alpha} \xrightarrow{\text{H}} \left(\begin{array}{c}
f_{i \alpha} \\
f_{i \alpha}^{\dagger}
\end{array}\right) \forall i,
&&
\mathcal{F}_{i\alpha} \xrightarrow{\text{U(1)}}\left(\begin{array}{c}
e^{i \theta_{i}} f_{i \alpha}^{\dagger} \\
e^{-i \theta_{i}} f_{i \alpha}
\end{array}\right),\nonumber \\ 
 \mathcal{F}_{i\alpha}\xrightarrow{\text{lPH}} \left(\begin{array}{c}
\alpha f_{i \bar{\alpha}} \\
\alpha f_{i \bar{\alpha}}^{\dagger}
\end{array}\right),
&&
\mathcal{F}_{i\alpha}\xrightarrow{\text{TR}} \left(\begin{array}{c}
e^{i\pi\alpha/2} f_{i \bar{\alpha}}^{\dagger} \\
e^{-i\pi\alpha/2} f_{i \bar{\alpha}}
\end{array}\right) \forall i. \label{eq:symDef}
\end{align}
Note, that H and TR also involve a complex conjugation when acting on complex numbers. We denote the spin index by $\alpha=\{\uparrow,\downarrow\}=\{+1,-1\}$ and $\bar{\alpha}$ indicates a spin-flip, $\bar{\alpha}=-\alpha$. 

The PF-term $\HPF$ is invariant under the local U(1) symmetry, but changes its sign under hermitian conjugation, global particle-hole (PH) and time-reversal symmetry,
\begin{align}
    \HS+\HPF \xrightarrow{\text{H},\text{PH},\text{TR}}  \HS-\HPF.
\end{align}
Therefore, the full Hamiltonian $\Bar{H}+\Bar{H}_\text{PF}$ is only symmetric under pairwise combinations of H, PH and TR symmetry. The  lPH symmetry ceases to be useful in the presence of $\HPF$.

Depending on the model-specific $J_{(i,j)}^{\mu\nu}$, we can further find lattice (L) or spin rotation (S) symmetries of $\HS$,
\begin{align}
    J^{\mu\nu}_{(i,j)} \xrightarrow{L} J^{\mu\nu}_{(i',j')}, && J^{\mu\nu}_{(i,j)} \xrightarrow{S} J^{\mu'\nu'}_{(i,j)}.
\end{align}
These symmetries are not broken by presence of $\HPF$.

\subsection{Symmetries of correlation functions}
The symmetries of the pf Hamiltonian impose symmetries on the correlation functions, which constitute the basic starting point for the fRG treatment. The single-particle correlation function (or propagator) is
\begin{equation}
G\left(1' ; 1\right)=\int_0^\beta d \tau' d \tau e^{i \tau' \omega'-i \tau \omega}\left\langle \mathcal{T}_\tau f_{i' \alpha'}^{\dagger}(\tau') f_{i \alpha}(\tau)\right\rangle,
\end{equation}
where we used the imaginary time-ordering and operators in the Heisenberg picture. The two-particle correlation function reads
\begin{multline}
G\left(1', 2' ; 1,2\right) = \\ \int_0^\beta d \tau_{1'} d \tau_{2'} d \tau_{1} d \tau_{2} e^{i\left(\tau_{1'} \omega_{1'}+\tau_{2'} \omega_{2'}-\tau_{1} \omega_{1}-\tau_{2} \omega_{2}\right)} \\
 \times\left\langle \mathcal{T}_\tau f_{i_{1'} \alpha_{1'}}^{\dagger}(\tau_{1'}) f_{i_{2'} \alpha_{2'}}^{\dagger}(\tau_{2'}) f_{i_{1}  \alpha_{1}}(\tau_{1}) f_{i_{2}  \alpha_{2}}(\tau_{2}) \right\rangle.
\end{multline}
On the left-hand side of the above equations, we use multi-indices,
$$
\begin{aligned}
1 & \equiv\left(i_{1}, \phantom{-}\omega_{1}, \alpha_{1}\right), \\
-1 & \equiv\left(i_{1},-\omega_{1}, \alpha_{1}\right), \\
\bar{1} & \equiv\left(i_{1}, \phantom{-}\omega_{1}, \bar{\alpha}_{1}\right).
\end{aligned}
$$
Following Ref.~\cite{buessen_functional_2019}, we summarize the symmetry constraints on the correlation functions in Tab.~\ref{tab:GreensFSymmetries1}.
For the two particle correlation function, there is the additional crossing symmetry (X) related to anticommuting two fermionic creation or annihilation operators.
\begin{align}
    &G\left(1', 2' ; 1,2\right) = -G\left(1', 2' ; 2,1\right) \\ \nonumber
    =\:& G\left( 2',1' ; 2,1\right) =  -G\left( 2',1' ; 1,2\right). 
\end{align}

\begin{table}[h]
    \caption{Symmetries of $\HS$ acting on $\HPF$ and correlation functions: Hermitian (H), local $\text{U}(1)$, global particle-hole (PH) and time-reversal (TR) symmetry.}
    \label{tab:GreensFSymmetries1}
    \centering
    \resizebox{\columnwidth}{!}{%
    \begin{tabular}{c|c|c|c}
          & $\HPF$ & $G\left(1' ; 1\right)$ & $G\left(1', 2' ; 1,2\right)$ \\\hline
        H  & $-\HPF$& $\phantom{-\alpha' \alpha}G(-1 ; -1')^*$ & $\phantom{\alpha_1\alpha_2\alpha_1'\alpha_2'}G(-1, -2 ; -1',-2')^*$ \\
       PH & $-\HPF$&$-\alpha' \alpha G(-\bar{1} ; -\bar{1}')\phantom{^*}$ & $\alpha_1\alpha_2\alpha_1'\alpha_2' G(-\bar{1}, -\bar{2} ; -\bar{1}',-\bar{2}')\phantom{^*}$    \\
       TR & $-\HPF$& $\phantom{-}\alpha' \alpha G(-\bar{1}' ; -\bar{1})^*$ & $\alpha_1\alpha_2\alpha_1'\alpha_2' G(-\bar{1}', -\bar{2}' ; -\bar{1},-\bar{2})^*$\\
    U(1)  & $\phantom{-}\HPF$ & $e^{i(\theta_{i_{1}'}-\theta_{i_{1}})}G\left(1' ; 1\right)$ & $e^{i(\theta_{i_{1}'}+\theta_{i_{2}'}-\theta_{i_{1}}-\theta_{i_{2}})}G\left(1', 2' ; 1,2\right)$
    \end{tabular}
    }
\end{table}

\subsection{Parameterization of correlators and vertices}
\label{subsec:parameterization}

The local U(1) symmetry \eqref{eq:symDef} constrains the single-particle correlator to be site-local and the two-particle correlator to be bi-local in real space. In addition, imaginary time translation symmetry reduces the number of independent frequencies by one. The dependence on the spin indices $\alpha$ can be parameterized by an expansion in Pauli matrices $\sigma^{\mu}$ with $\mu=0,x,y,z$ and $\sigma^0$ the identity matrix. These considerations allow for the parametrization \cite{buessen_functional_2019}
\begin{align}
G\left(1' ; 1\right)= \delta_{i' i} \delta_{\omega' , \omega} \sum_{\mu=0,x,y,z} G^{\mu}_i(\omega) \sigma_{\alpha' \alpha}^{\mu}, \label{eq:PropParam}
\end{align}
and
\begin{align}
G(1',2' ; 1,2) = &\label{eq:FourPointParam}\\
\times [ \big( \sum_{\mu,\nu=0,x,y,z} G_{i_{1} i_{2}}^{\mu \nu}&(s, t, u) \sigma_{\alpha_{1'} \alpha_{1}}^{\mu} \sigma_{\alpha_{2'} \alpha_{2}}^{v}\big) \delta_{i_{1'} i_{1}} \delta_{i_{2'} i_{2}}\nonumber\\
&-(1' \leftrightarrow 2' ) ]\delta_{\omega_{1'}+\omega_{2'},\,\omega_{1}+\omega_{2}}, \nonumber
\end{align}
where the bosonic transfer frequencies are
\begin{align}
&s=\omega_{1'}+\omega_{2'}= \omega_1+\omega_2,\\
&t=\omega_{1'}-\omega_{1}= \omega_2-\omega_{2'}, \\
&u=\omega_{1'}-\omega_{2}= \omega_1-\omega_{2'}.
\end{align}
The complex numbers $G^{\mu}_i(\omega)$ and $G_{i_{1} i_{2}}^{\mu \nu}(s,t,u)$ can be further constrained using the relations in the first three lines of Tab.~\ref{tab:GreensFSymmetries1}. This is summarized in Tab.~\ref{tab:GreensFSymmetries2}. 
\begin{table}[h]
 \caption{Constraints on the parameterized pf correlation functions of Eqns.~\eqref{eq:PropParam} and \eqref{eq:FourPointParam}. The constraints are labeled by the symmetries that have been used.}
    \label{tab:GreensFSymmetries2}
    \centering
    \begin{tabular}{@{}lr@{}}
    \hline
 $G_i^{\mu}(\omega)=\xi(\mu) G_i^{\mu}(\omega)$ & $(\mathrm{H} \circ \mathrm{TR})$ \\
$G_i^{\mu}(\omega)=-G_i^{\mu}(\omega)^{*}$ & $(\mathrm{TR} \circ \mathrm{PH})$ \\
\hline 
$G_{i_{1} i_{2}}^{\mu\nu}(s, t, u)=\xi(\mu) \xi(v) G_{i_{1} i_{2}}^{\mu \nu}(s,-t, u)$ & $(\mathrm{H} \circ \mathrm{TR}) $\\
$G_{i_{1} i_{2}}^{\mu\nu}(s, t, u)=\xi(\mu) \xi(v) G_{i_{2} i_{1}}^{v \mu}(s, t,-u)$ & $(\mathrm{X} \circ \mathrm{H} \circ \mathrm{TR})$ \\
$G_{i_{1} i_{2}}^{\mu\nu}(s, t, u)=\xi(\mu) \xi(v) G_{i_{1} i_{2}}^{\mu \nu}(s, t, u)^{*}$ & $( \mathrm{H} \circ \mathrm{PH})$ \\
\hline
    \end{tabular}
\end{table}\\
In these relations, we have introduced the sign function
\begin{align}
\xi(\mu)= \begin{cases}+1 & \text { if } \mu=0, \\ -1 & \text { otherwise. }\end{cases}
\end{align}
Our results indicate that the propagator takes the simple diagonal form
\begin{equation}
G(1',1)=\delta_{i',i}\delta_{\omega',\omega}\delta_{\alpha',\alpha} G_i(\omega) \label{eq:G1p1}
\end{equation}
and we define the real self-energy $\gamma_i(\omega)$ via
\begin{equation}
G_{i}(\omega)=\frac{1}{i\omega+i\gamma_{i}(\omega)}\equiv -ig_i(\omega) \in i\mathbb{R}. \label{eq:propagator}
\end{equation}
In comparison to the standard pf-fRG without PF potential term \cite{buessen_functional_2019}, $\gamma_i(\omega)$ is no longer anti-symmetric in $\omega$. Likewise, the two-particle correlator has no symmetry relating $s \leftrightarrow -s$ or $s \leftrightarrow u$. This reduction of symmetries amounts to a factor $\approx 4$ in memory and computation time compared to the standard pf-fRG scheme. 

Finally, the fRG flow equations in vertex expansion are written in terms of the (one-particle irreducible) vertices defined from the connected correlators via the tree expansion \cite{kopietz_introduction_2010}. According to Eq.~\eqref{eq:propagator}, we have for the self-energy $\Sigma(1';1)=\delta_{i',i}\delta_{\omega',\omega}\delta_{\alpha',\alpha}\left\{ -i\gamma_{i}(\omega)\right\}$. The two-particle vertex $\Gamma(1',2' ; 1,2)$ is defined from the connected part of $-G(1',2';1,2)$ by amputating external propagators. As the latter take the simple diagonal form \eqref{eq:G1p1}, the vertex can be parameterized in analogy to the correlator, i.e.~Eq.~\eqref{eq:FourPointParam} with $G$ replaced by $\Gamma$
and the symmetries listed in Tab.~\ref{tab:GreensFSymmetries2} are also applicable to $\Gamma_{i_1i_2}^{\mu\nu}(s,t,u)$.

So far, we have made no assumptions on the form of the spin-spin interaction $J_{(i,j)}^{\mu\nu}$. If present, spin rotation and lattice symmetries can be used to relate different site and spin indices of $\gamma_i(\omega)$ and $G_{i_{1} i_{2}}^{\mu \nu}$. From now on we focus on the SO(3) symmetric Heisenberg case,
\begin{equation}
    J_{(i,j)}^{\mu\nu} = \delta_{\mu,\nu} J_{(i,j)} , \label{eq:Heisenberg}
\end{equation}
for $\mu,\nu=x,y,z$ so that $G_{i_{1} i_{2}}^{00} \equiv G_{i_{1} i_{2}}^d \in  \mathbb{R}$ and $G_{i_{1} i_{2}}^{xx} = G_{i_{1} i_{2}}^{yy} = G_{i_{1} i_{2}}^{zz} \equiv G_{i_{1} i_{2}}^s \in  \mathbb{R}$ are the only non-vanishing correlation functions. Analogously, $\Gamma_{i_{1} i_{2}}^{s,d}$ are the only non-vanishing vertices.

\subsection{ppf-fRG}

For the (p)pf-fRG, a Matsubara cutoff scheme is applied to the bare propagator $G^{(0)}_j(\omega)=1/i\omega \rightarrow G^{(0),\Lambda}_j(\omega) \equiv \theta^\Lambda(\omega) G^{(0)}_j(\omega)$ \cite{reuther_j_2010,reuther_frustrated_2011} with cutoff function $\theta^\Lambda(\omega)$ smoothly interpolating from unity to zero as the magnitude of $\omega$ drops below the cutoff scale $\Lambda$. When $\Lambda=0$ the bare propagator, $G^{(0),\Lambda=0}_j(\omega)=1/i\omega$, is recovered and the action describes the physical system of interest. At $\Lambda=\infty$, however, the modified propagator vanishes and the vertex functions are trivial and frequency independent,
\begin{align}
   \gamma_{i}^{\Lambda=\infty}(\omega) =& \frac{\pi}{2\beta}, \label{eq:InitialCondition1}\\
    \Gamma_{i_1 i_2}^{s,\Lambda=\infty}(s,t,u) =& J_{(i_1,i_2)}/4, \\ 
     \Gamma_{i_1 i_2}^{d,\Lambda=\infty}(s,t,u) =& 0. \label{eq:InitialCondition3}
\end{align}
The Wetterich equation \cite{wetterich_exact_1993} describes the flow of all $n-$particle vertex functions under variation of $\Lambda$ from the trivial starting point $\Lambda=\infty$ to the physical endpoint $\Lambda = 0$. The resulting hierarchy of flow equations can usually not be solved exactly, but has to be truncated, with multiple truncation schemes available. Here, we focus on the established one-loop scheme \cite{kopietz_introduction_2010} together with Katanin truncation \cite{katanin_fulfillment_2004} which partially considers the effect of the three-particle vertex in the flow of the two-particle vertex and constitutes the standard choice in the pf-fRG literature \cite{reuther_j_2010}.
We refer to App.~\ref{App:FlowEq} for the choice of cutoff function and the flow equations which do not differ from the standard pf-fRG case. We use discrete Matsubara grids for all frequency arguments with about $N_w=30$ (positive) frequencies and ensure that our results are converged in $N_w$. The ppf-fRG flows are smooth in $\Lambda$ with features appearing around $\Lambda \sim J$ and a plateau towards $\Lambda \rightarrow 0$ from which we obtain the end-of-flow results reported in the following.  In App.~\ref{app:Observables}, we show how imaginary frequency spin susceptibilities $\chi_{i_1i_2}(\Omega)$ and equal-time spin correlation functions like the spin projection $C_i(T)$ are computed, with a technical subtlety appearing for the bubble contribution in the latter case. In summary, the main difference between the ppf-fRG and the pf-fRG is the slightly reduced symmetry of the vertex functions discussed in the previous subsection, the finite initial condition for the self energy and fRG flows that are smooth and convergent.

\subsection{ppf-PA}

Recently, a lot of effort has been put into generalizing the one-loop truncation of general fermionic fRG flows to higher loop orders using multiloop fRG (mfRG) \cite{kugler_multiloop_2018,kugler_multiloop_2018-1,kugler_derivation_2018}. 
Applications include the Anderson impurity model \cite{chalupa-gantner_fulfillment_2022} and the two-dimensional Hubbard model \cite{tagliavini_multiloop_2019,hille_quantitative_2020}.
By construction, vertices obtained with mfRG in the limit of infinite loops converge to solutions of the parquet approximation (PA), a complementary diagrammatic formulation of the many-body problem in which the self-energy and frequency-dependent contributions to the two-particle vertex are self-consistently described by the Schwinger-Dyson and three Bethe-Salpeter equations, respectively. 
In-depth discussions of their structure for pseudo-fermion Hamiltonians can, for example, be found in Refs.~\cite{thoenniss_multiloop_nodate, kiese_multiloop_2022}. In contrast to the full parquet equations, the PA neglects frequency dependent contributions to the fully two-particle irreducible vertex $I_{\text{2PI}}$ which thus reduces to the bare vertex $\Gamma_0$, with $\Gamma_0 \sim J$ for pf systems. Deviations from exact vertices set in at fourth order in $J / T$, corresponding to the so-called \textit{envelope} diagram.

Using the very same initial conditions as in the $\Lambda \to \infty$ limit of ppf-fRG \eqref{eq:InitialCondition1}--\eqref{eq:InitialCondition3}, we numerically converge the algebraic equations of the PA using forward iterations combined with a mixed update scheme, which determines the input for the next iteration as
\begin{align}
    x_{\text{new}} = (1 - \lambda) x_{\text{old}} + \lambda f(x_{\text{old}}) \,,
\end{align}
where $x = (\Sigma, \Gamma)$ and $f$ schematically denotes the Schwinger-Dyson and Bethe-Salpeter equations. For $J / T \ll 1$, full updates ($\lambda = 1$) were sufficient to meet the convergence criterion $|| f(x_{\text{old}}) - x_{\text{old}} || < 10^{-6}$ between subsequent iterations ($|| . ||$  is the maximum norm). For lower temperatures $\lambda$ had to be reduced from unity in order to obtain converged results. We choose extended Matsubara grids with up to $48 \cross 24^2$ frequencies in mixed bosonic-fermionic frequency notation to parameterize the two-particle vertex as well as $32$ frequencies for the self-energy. For the vertex, we take into account the decomposition of each channel into its respective asymptotic functions $K_1$, $K_2$ ($\bar{K}_2$) and $R$ as detailed in Ref.~\cite{wentzell_high-frequency_2020}. We checked convergence with respect to the number of frequencies and, in most cases (see App.~\ref{App:TemperatureScaling} and \ref{app:PhaseTransitions}), found no significant changes of our results if more were included.


\section{Results}
\label{sec:Results}

\subsection{Benchmark: Small spin clusters}
\label{subsec:benchmark}

\begin{figure}[t]
    \centering
    \includegraphics{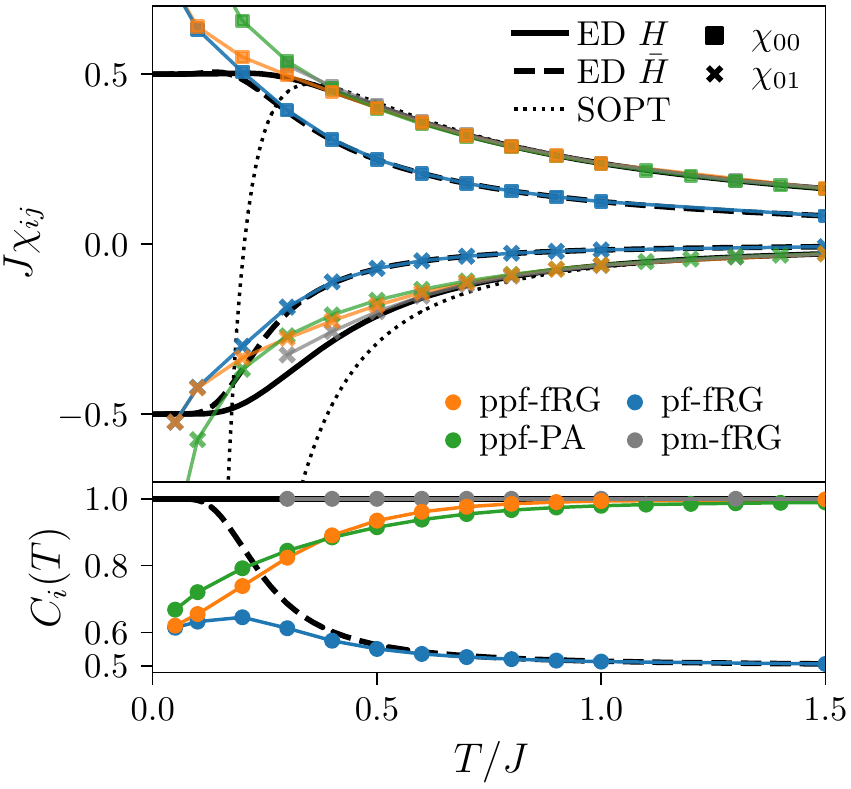}
    \caption{Heisenberg dimer: Local and non-local susceptibilities (top panel: squares and crosses, respectively) and spin projection $C_i(T)=4\langle \Bar{S}^z_i \Bar{S}^z_i \rangle$ (bottom panel) computed with PF projection using the fRG (ppf-fRG) and the parquet approximation (ppf-PA). For comparison, we show the exact results (solid lines), second-order perturbation theory in $J$ (dotted lines), pf-fRG results without PF projection and data from the pseudo-Majorana fRG (pm-fRG) building on a fermionic spin representation without unphysical states.
    } 
    \label{fig:DimerBenchmark}
\end{figure}

To benchmark the proposed ppf-fRG and ppf-PA approaches, we consider the AFM Heisenberg dimer $H_\mathrm{dimer}=J\mathbf{S}_0\cdot\mathbf{S}_1$ with $J=1$ and focus on the static local and non-local susceptibilities and the spin projection $C_i$ as a function of $T$. Exact ED results of the spin Hamiltonian are shown as solid black lines in Fig.~\ref{fig:DimerBenchmark}, see Ref.~\cite{niggemann_frustrated_2021} for closed-form expressions. The (end-of-flow) ppf-fRG results are denoted by orange symbols. For large and moderate temperatures $T \gtrsim 0.4 J$, the exact susceptibilities are accurately reproduced by the ppf-fRG. For lower $T$, the susceptibilities become unphysically large in magnitude. This breakdown of accuracy is also reflected in $C_i$ which considerably drops below unity with decreasing $T$. 

The inaccuracies at low $T$ are due to the truncation of the ppf-fRG flow equations so that self-energy and (two-particle) vertex are only correct up to order $J^2/T$, see App.~\ref{App:TemperatureScaling} for numerical confirmation. However, the advantage of fRG over naive second-order perturbation theory (SOPT, dotted line in Fig.~\ref{fig:DimerBenchmark}) is the re-summation of certain diagrams to infinite order \cite{reuther_frustrated_2011}, which is essential for the detection of magnetic ordering tendencies (see below). In the dimer case, the re-summation stabilizes the susceptibilities beyond temperatures where SOPT is applicable.

As the treatment of the full three-particle vertex is prohibitively expensive, an interesting question is if the above truncation problem can be alleviated by invoking higher-loop orders beyond the Katanin truncation. This has been shown to be numerically feasible in the pf-fRG \cite{kiese_multiloop_2022,thoenniss_multiloop_nodate}. In fact, as explained in Sec.~\ref{sec:correlators}, our finite-temperature application makes it even possible to converge the ppf-PA equations equivalent to the loop-converged multi-loop result with an error of order $J^4/T^3$ and $J^5/T^4$ for vertex and self-energy, respectively. Surprisingly, our ppf-PA results (green symbols in Fig.~\ref{fig:DimerBenchmark}) show no systematic improvement compared to the ppf-fRG. This unexpected finding also questions the usefulness of multi-loop extensions in the context of pf applications.

For completeness, blue symbols in Fig.~\ref{fig:DimerBenchmark} show results of the standard one-loop pf-fRG applied to $\Bar{H}$ (without PF potential $\HPF$) which at large $T$ compare well with the exact (but unphysical) ED results obtained from $\Bar{H}$ (dashed black line). Like in the ppf-fRG, agreement only holds for $T \gtrsim 0.4 J$, suggesting the truncation error in the fRG flow equations is largely independent of $\HPF$. 

\begin{figure}[t]
    \centering
    \includegraphics{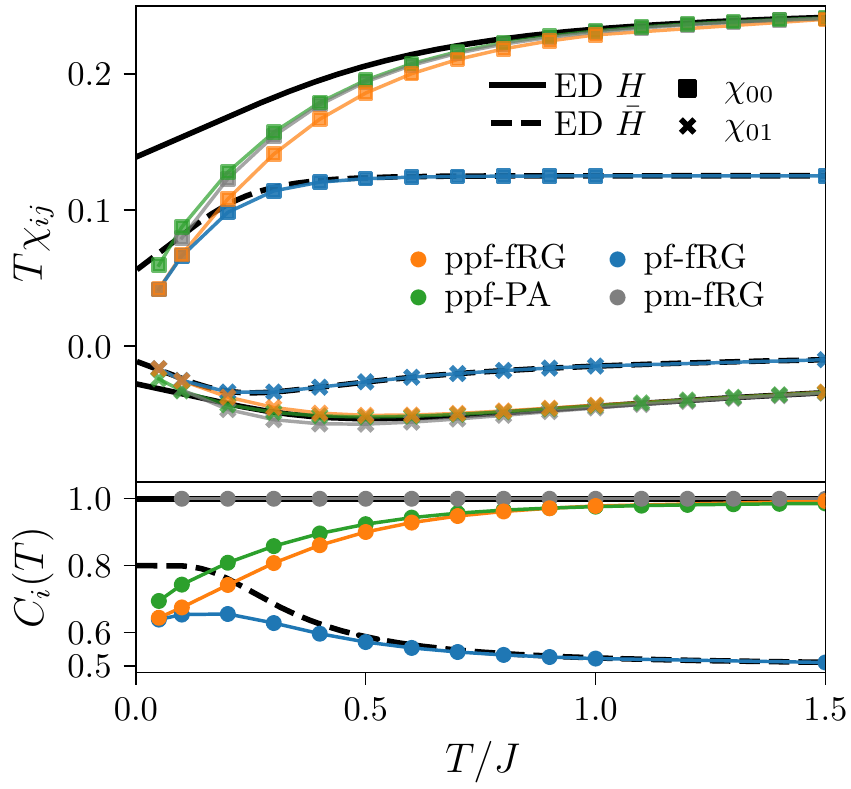}
    \caption{Heisenberg trimer: The top panel shows the static susceptibilities that diverge as $T\rightarrow 0$ and have been multiplied by temperature to obtain finite values at $T=0$. The bottom panel shows the spin projection $C_i(T)$. The symbols follow the convention of Fig.~\ref{fig:DimerBenchmark}.
    }
    \label{fig:TrimerBenchmark}
\end{figure}

Finally, in Fig.~\ref{fig:TrimerBenchmark} we consider the Heisenberg trimer, Eq.~\eqref{eq:trimer}, where the ppf-fRG and ppf-PA results are qualitatively similar to the dimer and strengthen the conclusions given above for the latter case. It is interesting to note, that, presumably due to the truncation of the flow equations, the ppf-fRG and pf-fRG converge to the same result for low $T$ even though the ground state of the pf trimer is not in the physical Hilbert space.

\subsection{Finite temperature magnetization transition in three dimensions}
\label{sec:PhaseTransitions}
One of the most remarkable properties of spin systems at finite $T$ is the possible appearance of a magnetization transition at a critical temperature $T_c$. As we consider (short-range coupled) Heisenberg systems with continuous SO(3) spin rotation symmetry, these transitions only occur at dimension three. Despite their classical nature, it is interesting to study magnetic phase transitions in models of -- possibly frustrated -- quantum spins. Here we investigate if ppf-fRG can detect the N\'{e}el transition in a nearest-neighbor cubic lattice Heisenberg AFM ($J_1=1$) using finite-size scaling of the correlation ratio \cite{sandvik_computational_2010,kaul_spin_2015,pujari_interaction-induced_2016},
\begin{align}
    \xi/L = \frac{1}{2\pi}\sqrt{\chi(\mathbf{Q})/\chi(\mathbf{Q}+\boldsymbol{\delta})-1},\label{eq:correlationRatio}
\end{align}
where $\chi(\mathbf{q})=\sum_j e^{-i\mathbf{q}\cdot(\mathbf{r}_i-\mathbf{r}_j)}\chi_{ij}$ is the momentum space static susceptibility, $\mathbf{Q}=(\pi,\pi,\pi)$ is the ordering wave vector, $\boldsymbol{\delta}=(0,0,2\pi/L)$ and $L$ is a measure of the system size. 

In the ppf-fRG of the cubic lattice AFM we assume a translation invariant infinite system, but limit the range of allowed non-trivial correlations by restricting the vertex functions $\Gamma^{s,d}_{i_1i_2}$ to $|\mathbf{r}_{i_1}-\mathbf{r}_{i_2}|\leq \tilde{L}$ \cite{reuther_j_2010}. The length scale $\tilde{L}$ can be used for finite-size scaling \cite{niggemann_quantitative_2022}. To smooth out discrete-lattice effects, we define $L=2(\frac{3}{4\pi} N)^{1/3}\simeq 2\tilde{L}$ to be used in Eq.~\eqref{eq:correlationRatio}, where $N$ is the number of sites to which the reference site is connected by a possible non-trivial vertex (including the on-site vertex). This particular choice of $L$ corresponds to the diameter of the smeared-out {\it correlation}-sphere including $N$ sites.

\begin{figure}[h!]
    \centering
    \includegraphics{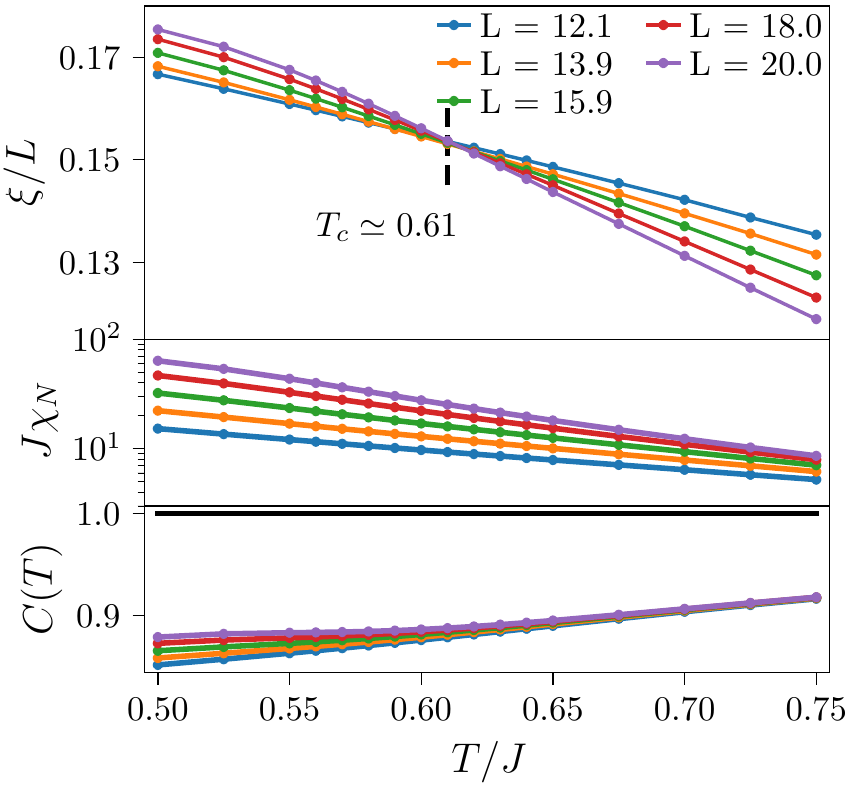}
    \caption{ppf-fRG: Correlation length $\xi$, N\'{e}el susceptibility $\chi_N$ and spin projection $C$ for the nearest neighbor AFM Heisenberg model on the cubic lattice. We find a line crossing at $T_c^\mathrm{fRG} \simeq 0.61$. The corresponding scaling collapse can be found in Fig.~\ref{fig:1lCubicAFMCollapse}. 
    }
    \label{fig:1lCubicAFM}
\end{figure}

\begin{figure}[h!]
    \centering
    \includegraphics{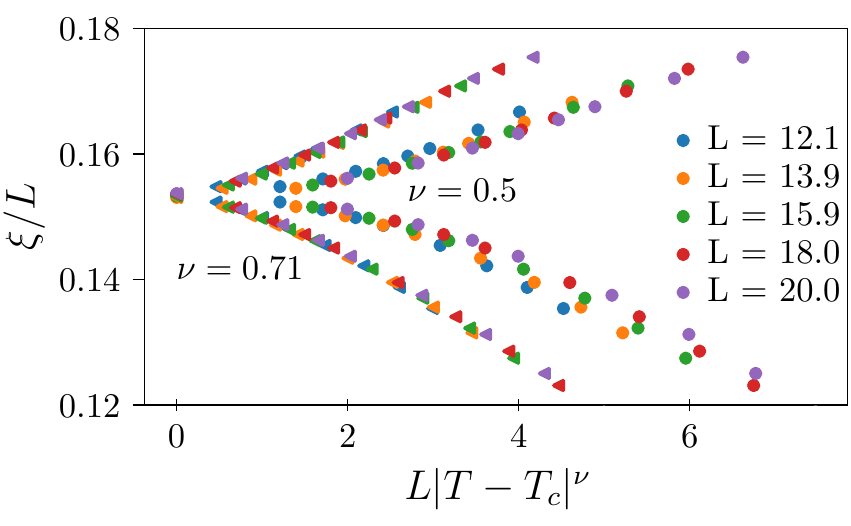}
    \caption{ppf-fRG: Collapse plot for the   data of Fig.~\ref{fig:1lCubicAFM}, assuming the mean-field $\nu=0.5$ (points) or exact $\nu\simeq0.71$ (triangles) correlation length critical exponents.
    }
    \label{fig:1lCubicAFMCollapse}
\end{figure}

Close to the critical temperature $T_c$, the anticipated scaling form of the AFM spin correlation length $\xi$ is
\begin{align}
    \xi/L \sim g_\pm(L\abs{T-T_c}^\nu),
\end{align}
so that $\xi/L$ becomes independent of $L$ at $T=T_c$, the sign $\pm$ refers to the sign of $T-T_c$ and $\nu$ is the universal critical exponent \cite{cardy_scaling_1996}. 
The ppf-fRG results for the correlation length, N\'{e}el susceptibility $\chi_N=\chi(\mathbf{Q})$ and the spin projection $C(T)$ are shown in Fig.~\ref{fig:1lCubicAFM}. We indeed find a clear line-crossing in the $\xi/L$ data indicating $T^{\text{fRG}}_c \simeq 0.61$ significantly below the error controlled quantum Monte-Carlo result $T_c^{\text{QMC}}=0.946(1)$ \cite{sandvik_critical_1998}. This might be related to an underestimation of the spin projection $C(T)$. The scaling collapse in Fig.~\ref{fig:1lCubicAFMCollapse} shows consistency with the correct three-dimensional Heisenberg universality class with $\nu \simeq 0.71$ \cite{chester_bootstrapping_2021-1}.
In App.~\ref{app:PhaseTransitions}, we further investigate the truncation dependence of these quantities considering analogous simulations using the ppf-fRG without Katanin truncation and in the random phase approximation (RPA) where both $C(T)$ and $T_c$ are overestimated and the critical exponent is consistent with the mean-field value $\nu= 0.5$. We also discuss results obtained from ppf-PA which indicates that this method is not well suited for the assessment of magnetic ordering transitions.

\section{Conclusion}
\label{sec:conclusion}

In summary, we have considered simple small spin clusters as examples to show that working with the pf representation \eqref{eq:pf} generally requires the PF projection to the physical $S=1/2$ subspace at all temperatures. This challenges the existing approach in the pf-fRG literature, which omits the projection and focuses on zero temperature. We have leveraged the pf-fRG to include the PF projection in potential form which requires only minor technical modifications and therefore can be readily adopted to existing variants of the method. The discrete nature of finite-temperature Matsubara frequencies simplifies the numerical implementation considerably. 

In conclusion, for small benchmark models the proposed ppf-fRG yields quantitatively reliable end-of-flow results at large and intermediate temperatures but fails at small temperatures $T \lesssim 0.4J$ due to the omission of higher-order in $J/T$ diagrams, a problem that also could not be alleviated by adopting a parquet scheme. We note that a detailed understanding for the surprisingly poor performance of the latter is lacking at this point. For forthcoming applications, we suggest that the deviation of the spin projection $C_i(T)$ from the exact value of unity by $10\%$ or more signals the quantitative failure of the ppf-fRG. 

This failure of the (p)pf-fRG at low and zero temperature also questions an earlier proposal to remove unphysical contributions of the pf representation in the context of the pf-fRG at $T=0$. The idea rests on the addition of an on-site term $- J_0 \sum_{\mu=x,y,z}\bar{S}^\mu_i\bar{S}^\mu_i$ \cite{baez_numerical_2017} with $J_0>0$ penalizing the $S=0$ sector energetically versus the $S=1/2$ sector. If the treatment of the pf Hamiltonian was exact, the above idea would be a valid $T=0$ alternative to the PF trick (the latter works at any $T$). However, due to the unavoidable approximations in the diagrammatic methods at hand, results obtained from the on-site term \cite{baez_numerical_2017,iqbal_quantum_2019} should be considered with appropriate care.

For infinite systems in three dimensions, we have demonstrated that it is possible to robustly detect magnetic ordering by finite-size scaling of the ppf-fRG data. This establishes an alternative to the less physical and implementation dependent concepts of flow-divergence or flow-breakdown at cutoff scale $\Lambda_\star$ which, by mean-field arguments, can be related to $T_c$ \cite{reuther_finite-temperature_2011,iqbal_functional_2016, iqbal_quantum_2019}.  However, although the correct type of N\'{e}el order has been predicted by the ppf-fRG for the nearest-neighbor AFM Heisenberg model on the cubic lattice, it remains unclear why the transition appears at a critical temperature more than $30\%$ below benchmark results. A possible hint might be the spin projection $C_i(T=T_c^\mathrm{fRG})\simeq 0.88$ which is out of the range identified as reliable above. 

Finally, there exist two alternatives to the PF projection scheme that we would like to mention. First, in a field theory framework, a functional delta-function representation can be used to constrain the pf operators to the physical subspace. This then leads to the notion of a bosonic gauge field which plays an important role in the (mean-field) theory of spin liquids \cite{wen_quantum_2007}. On the computational side, however, the introduction of such a bosonic field would require a multitude of additional vertex functions \cite{kopietz_introduction_2010}, a formidable challenge yet to be faced.
Second, it is well known that a faithful quantum spin $S=1/2$ representation exists in terms of Majorana fermions \cite{martin_generalized_1959,tsvelik_new_1992}. The associated pseudo-Majorana fRG (pm-fRG) has been developed only recently \cite{niggemann_frustrated_2021}. The data from pm-fRG applied to the benchmark clusters treated in Sec.~\ref{subsec:benchmark} are shown by the gray lines in Fig.~\ref{fig:DimerBenchmark} (dimer) and Fig.~\ref{fig:TrimerBenchmark} (trimer). The pm-fRG results are very similar to those of the ppf-fRG, with the same difficulties at small $T$. However, for the detection of the magnetic phase transition in the cubic lattice case, the pm-fRG determines $T_c$ much more accurately only $5\%$ below the exact result \cite{niggemann_quantitative_2022}.

\section*{Acknowledgements}
 We thank Johannes Reuther, Nils Niggemann, Jan von Delft, Ronny Thomale and Marc Ritter for fruitful discussions. The authors gratefully acknowledge the Gauss Centre for Supercomputing e.V. (www.gauss-centre.eu) for funding this project by providing computing time through the John von Neumann Institute for Computing (NIC) on the GCS Supercomputer JUWELS at Jülich Supercomputing Centre (JSC). The computations in this work were, in part, run at facilities supported by the Scientific Computing Core at the Flatiron Institute, a division of the Simons Foundation. We acknowledge the usage of the Noctua2 cluster at the Paderborn Center for Parallel Computing (PC$^2$). The authors acknowledge financial support by a MCQST-START fellowship and by the Munich Quantum Valley, which is supported by the Bavarian state government with funds from the Hightech Agenda Bayern Plus. 

\appendix
\begin{widetext}

\section{ppf-fRG flow equations for the Heisenberg case}
\label{App:FlowEq}
The one-loop flow equations for the pf-fRG as found in Refs.~\cite{reuther_frustrated_2011,hering_new_2019} do not rely on the vertex symmetries that are broken by the PF potential term \eqref{eq:PopovPotential}, see the discussion in Sec.~\ref{sec:correlators}. Here, we write these flow equations for the Heisenberg case where the vertices are parameterized in a density and spin part, $\Gamma^s$ and $\Gamma^d$, see Sec.~\ref{subsec:parameterization}. \\
The cutoff dependent propagator is defined as $ G^\Lambda_j(\omega) = -i g^\Lambda_j(\omega) =\frac{\theta^\Lambda(\omega)}{i\omega+i\theta^\Lambda(\omega)\gamma_j(\omega)}$ where $\Sigma_j(\omega)=-i\gamma_j(\omega)$ and we chose the regulator to be Lorentzian $\theta^\Lambda(\omega)= \frac{\omega^2}{\omega^2+\Lambda^2}$.
The flow equation for the spin part of the vertex reads 
\begin{align}
\frac{d}{d\Lambda}\Gamma^{s,\Lambda}_{i_{1} i_{2}}(s,t,u)&= T \sum_{\omega}\label{eq:dGamma_s}\\ P^\Lambda_{i_1i_2}(\omega,s-\omega)\times&[-2\Gamma^{s,\Lambda}\Gamma^{s,\Lambda}+\Gamma^{d,\Lambda}\Gamma^{s,\Lambda}+\Gamma^{s,\Lambda}\Gamma^{d,\Lambda}](s,\omega_{1'}-\omega,\omega-\omega_{2'})_{i_1i_2}(s,\omega-\omega_{1},\omega-\omega_{2})_{i_1i_2}\nonumber\\
-\sum_j P^\Lambda_{jj}(\omega,\omega-t)\times&[2\Gamma^{s,\Lambda}\Gamma^{s,\Lambda}](\omega_1+\omega,t,\omega_{1'}-\omega)_{i_1j}(\omega+\omega_{2'},t,\omega-\omega_2)_{ji_2}\nonumber\\
+P^\Lambda_{i_2i_2}(\omega,\omega-t)\times&[\Gamma^{s,\Lambda}\Gamma^{d,\Lambda}-\Gamma^{s,\Lambda}\Gamma^{s,\Lambda}](\omega_1+\omega,t,\omega_{1'}-\omega)_{i_1i_2}(\omega+\omega_{2'},\omega-\omega_2,t)_{i_2i_2}\nonumber\\
+P^\Lambda_{i_1i_1}(\omega,\omega-t)\times&[\Gamma^{d,\Lambda}\Gamma^{s,\Lambda}-\Gamma^{s,\Lambda}\Gamma^{s,\Lambda}](\omega_1+\omega,\omega_{1'}-\omega,t)_{i_1i_1}(\omega+\omega_{2'},t,\omega-\omega_2)_{i_1i_2}\nonumber\\
+P^\Lambda_{i_1i_2}(\omega,\omega+u)\times&[2\Gamma^{s,\Lambda}\Gamma^{s,\Lambda}+\Gamma^{d,\Lambda}\Gamma^{s,\Lambda}+\Gamma^{s,\Lambda}\Gamma^{d,\Lambda}](\omega_{1'}+\omega,\omega_{2}-\omega,u)_{i_1i_2}(\omega+\omega_{1},\omega-\omega_{2'},u)_{i_1i_2}\nonumber
\end{align}
and the density part flows according to
\begin{align}
\frac{d}{d\Lambda}\Gamma^{d,\Lambda}_{i_{1} i_{2}}(s,t,u)&= T \sum_{\omega}\label{eq:dGamma_d}\\ P^\Lambda_{i_1i_2}(\omega,s-\omega)\times&[\Gamma^{d,\Lambda}\Gamma^{d,\Lambda}+3\Gamma^{s,\Lambda}\Gamma^{s,\Lambda}](s,\omega_{1'}-\omega,\omega-\omega_{2'})_{i_1i_2}(s,\omega-\omega_{1},\omega-\omega_{2})_{i_1i_2}\nonumber\\
-\sum_j P^\Lambda_{jj}(\omega,\omega-t)\times&[2\Gamma^{d,\Lambda}\Gamma^{d,\Lambda}](\omega_1+\omega,t,\omega_{1'}-\omega)_{i_1j}(\omega+\omega_{2'},t,\omega-\omega_2)_{ji_2}\nonumber\\
+P^\Lambda_{i_2i_2}(\omega,\omega-t)\times&[3\Gamma^{d,\Lambda}\Gamma^{s,\Lambda}+\Gamma^{d,\Lambda}\Gamma^{d,\Lambda}](\omega_1+\omega,t,\omega_{1'}-\omega)_{i_1i_2}(\omega+\omega_{2'},\omega-\omega_2,t)_{i_2i_2}\nonumber\\
+P^\Lambda_{i_1i_1}(\omega,\omega-t)\times&[3\Gamma^{s,\Lambda}\Gamma^{d,\Lambda}+\Gamma^{d,\Lambda}\Gamma^{d,\Lambda}](\omega_1+\omega,\omega_{1'}-\omega,t)_{i_1i_1}(\omega+\omega_{2'},t,\omega-\omega_2)_{i_1i_2}\nonumber\\
+P^\Lambda_{i_1i_2}(\omega,\omega+u)\times&[\Gamma^{d,\Lambda}\Gamma^{d,\Lambda}+3\Gamma^{s,\Lambda}\Gamma^{s,\Lambda}](\omega_{1'}+\omega,\omega_2-\omega,u)_{i_1i_2}(\omega+\omega_{1},\omega-\omega_{2'},u)_{i_1 i_2}\nonumber
\end{align}
where
\begin{equation}
    P^\Lambda_{ij}(\omega,\omega')= \left(-ig^\Lambda_i(\omega)\right)S^\Lambda_j(\omega') + \left(-ig_j^\Lambda(\omega')\right)S^\Lambda_i(\omega),
\end{equation}
and $S^\Lambda_i(\omega)$ is the single scale propagator
\begin{align}
    S^\Lambda_j(\omega) =  -\frac{\partial}{\partial\Lambda}G^\Lambda_j(\omega)= \left(-ig_j^\Lambda(\omega)\right)\frac{\partial}{\partial\Lambda}\frac{i\omega}{\theta^\lambda(\omega)}\left(-ig_j^\Lambda(\omega)\right)= -i \left(g_j^\Lambda(\omega)\right)^2\frac{\partial}{\partial\Lambda}\frac{\omega}{\theta^\lambda(\omega)}.
\end{align}
Finally, the flow equation for the self energy is
\begin{align}
    \frac{d}{d \Lambda}\gamma^\Lambda_i(\omega_1) = T\sum_{\omega_2} \sum_j [2\Gamma_d(\omega_1+\omega_2,0,\omega_1-\omega_2)_{ij}-\delta_{ij}(3\Gamma_s+\Gamma_d)(\omega_1+\omega_2,\omega_1-\omega_2,0)_{ii}]S_j(\omega_2). \label{eq:SelfEnergyFlow}
\end{align}
In the Katanin truncation scheme, the partial derivative in the single scale propagator becomes a full derivative only in the flow equations for the vertex \cite{reuther_frustrated_2011},
\begin{align}
    S^\Lambda_j(\omega) =  -\frac{\dd}{\dd\Lambda}G^\Lambda_j(\omega)=  -i (g_j^\Lambda(\omega))^2\left(\frac{\partial}{\partial\Lambda}\frac{\omega}{\theta^\lambda(\omega)}+\frac{\dd}{\dd\Lambda}\gamma_j^\Lambda(\omega)\right).
\end{align}
\section{Observables from vertex functions}
\label{app:Observables}

Following Ref.~\cite{thoenniss_multiloop_nodate} the susceptibilities $\chi_{i j}(\Omega)\equiv\chi_{i j}^{zz}(\Omega)$ for the Heisenberg case can be computed from the self energy and vertices
\begin{align}
    \chi_{i j}^{zz}(\Omega)
=&\int_{0}^{\beta} \mathrm{d} \tau e^{i \Omega \tau} \sum_{\alpha_{1},\alpha_{1'},\alpha_{2} ,\alpha_{2'}} 
\frac{1}{4} \sigma_{\alpha_{1} \alpha_{1'}}^{z} \sigma_{\alpha_{2} \alpha_{2'}}^{z}
\langle\mathcal{T}_{\tau} f^\dagger_{i \alpha_{1}}(\tau) f_{i \alpha_{1'}}(\tau) f^\dagger_{j \alpha_{2}}(0) f_{j \alpha_{2'}}(0)\rangle\\
=&-\frac{\delta_{ij}}{2\beta}\sum_\omega G_i(\omega)G_i(\omega+\Omega)\\
& -\frac{\delta_{ij}}{2\beta^2}\sum_{\omega,\omega'} G_i(\omega)G_i(\omega+\Omega)G_i(\omega')G_i(\omega'+\Omega)\times[\Gamma^{s,\Lambda}-\Gamma^{d,\Lambda}](\omega+\omega'+\Omega,\omega-\omega',\Omega)_{ii}\nonumber\\
& -\frac{1}{2\beta^2}\sum_{\omega,\omega'} G_i(\omega)G_i(\omega+\Omega)G_j(\omega')G_j(\omega'+\Omega)\times 2\Gamma^{s,\Lambda}(\omega+\omega'+\Omega,\Omega,\omega-\omega')_{ij}.\nonumber
\end{align}
To compute the equal time susceptibility we have to sum over all bosonic Matsubara frequencies $\Omega$ and use the infinitesimal positive imaginary time $\delta\tau$,
\begin{align}
    \langle S^z_iS^z_j\rangle
=&  \sum_{\alpha_{1},\alpha_{1'},\alpha_{2},\alpha_{2'}} 
\frac{1}{4} \sigma_{\alpha_{1} \alpha_{1'}}^{z} \sigma_{\alpha_{2} \alpha_{2'}}^{z}
\langle\mathcal{T}_{\tau} f^\dagger_{i \alpha_{1}}(+\delta\tau) f_{i \alpha_{1'}}(+\delta\tau) f^\dagger_{j \alpha_{2}}(0) f_{j \alpha_{2'}}(0)\rangle\\
=&-\frac{ \delta_{ij}}{2\beta^2} \lim_{\delta\tau\xrightarrow[]{}0}[\sum_\omega e^{i\delta\tau\omega} G_i(\omega)][\sum_\omega e^{-i\delta\tau\omega} G_i(\omega)]\\
& -\frac{\delta_{ij}}{2\beta^3}\sum_{\omega,\omega',\Omega} G_i(\omega)G_i(\omega+\Omega)G_i(\omega')G_i(\omega'+\Omega)\times[\Gamma^s-\Gamma^d](\omega+\omega'+\Omega,\omega-\omega',\Omega)_{ii}\nonumber\\
& -\frac{1}{2\beta^3}\sum_{\omega,\omega',\Omega} G_i(\omega)G_i(\omega+\Omega)G_j(\omega')G_j(\omega'+\Omega)\times2\Gamma^s(\omega+\omega'+\Omega,\Omega,\omega-\omega')_{ij}.\nonumber
\end{align}
In the bubble term, the limit $\delta\tau\xrightarrow[]{}0$ has to be taken with care and can not be straightforwardly computed numerically. We compute it by adding and subtracting the sum over the bare part of the propagator and calculate the second sum analytically. The remaining sum can be calculated numerically.
\begin{align}
    \lim_{\delta\tau\xrightarrow[]{}0}\sum_\omega e^{i\delta\tau\omega} G(\omega) &= \underbrace{\sum_\omega (G(\omega) +\frac{i}{\omega+T\frac{\pi}{2}})}_{\text{finite}}-\lim_{\delta\tau\xrightarrow[]{}0}\sum_\omega e^{i\delta\tau\omega} \frac{i}{\omega+\frac{\pi}{2\beta}}\\
&= \sum_\omega (G(\omega)\nonumber +\frac{i}{\omega+\frac{\pi}{2\beta}})+i\beta(\frac{1}{2}-\frac{i}{2}).
\end{align}

\section{Perturbative check for ppf-fRG and ppf-PA at large temperatures}
\label{App:TemperatureScaling}

When considering large or intermediate temperatures, the one-loop ppf-fRG and the ppf-PA are error controlled
with respect to the exact solution. In the one-loop truncation, diagrams of order $\frac{J^3}{T^2}$ and higher are neglected when
calculating self energy or vertex. Therefore, the difference of the ED solution with the fRG solution should scale with
$\frac{J^3}{T^2}$. This behavior can be seen in Fig.~\ref{fig:TemperatureScaling}(a) for the self-energy of the Heisenberg dimer. For small temperatures,
when $\frac{J}{T}$ becomes large, the scaling breaks down. In the ppf-PA, diagrams of order $\frac{J^5}{T^4}$ are neglected for the self
energy. Compared to one-loop, this error scaling in ppf-PA is challenging to observe and can only be seen for very
large frequency grids and at high temperatures, see Fig.~\ref{fig:TemperatureScaling}(b).

\begin{figure}[h]
    \centering
    \subfloat{
\includegraphics{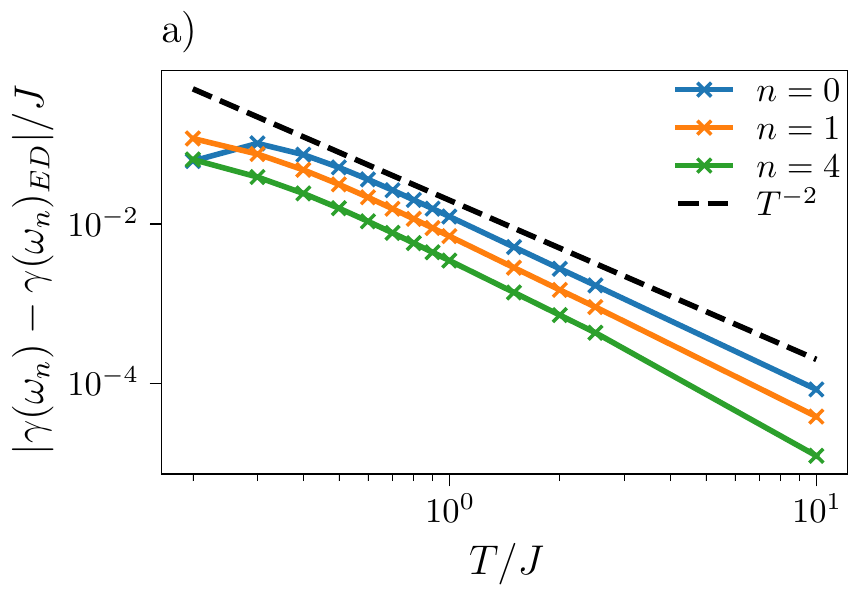}
    }\hfill
    \subfloat{
\includegraphics{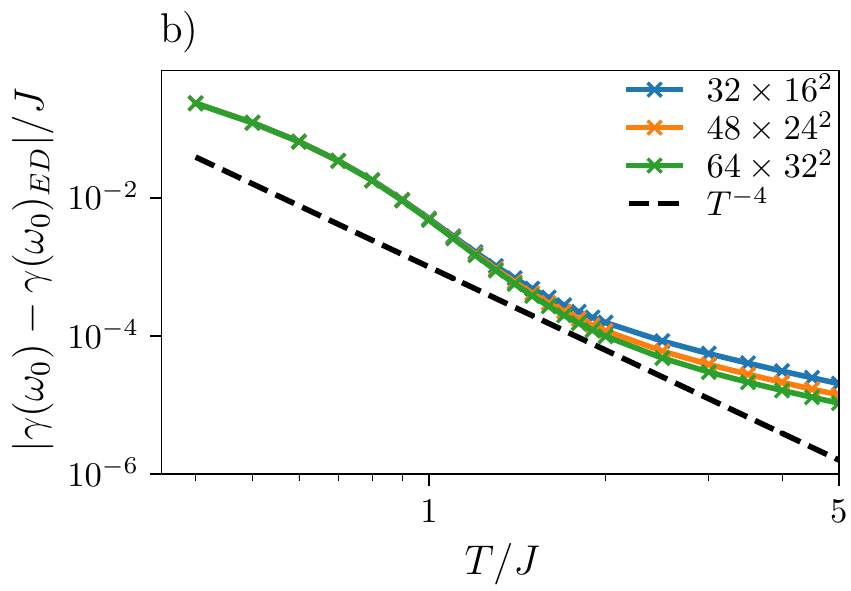}
}\\

 \caption{Error of the pf self energy in the Heisenberg dimer with $J=1$: a) In ppf-fRG, the error scales with $\frac{J^3}{T^2}$. For small temperatures, this scaling breaks down because $J/T$ ceases to be a small parameter. b) In ppf-PA, the observed error scaling depends on the the frequency grid, three choices are indicated by different colors. For larger frequency grids the self-energy error approaches the expected scaling with $\frac{J^3}{T^4}$ more closely. }
    \label{fig:TemperatureScaling}
\end{figure}

\section{Finite temperature phase transitions for other truncation schemes}
\label{app:PhaseTransitions}

In Sec.~\ref{sec:PhaseTransitions} the ppf-fRG was applied to asess the magnetic phase transition in the cubic lattice AFM Heisenberg model. Here, we additionally investigate the ppf-fRG without Katanin truncation, the RPA \cite{reuther_j_2010, reuther_frustrated_2011}, and the ppf-PA as further benchmark. The numerical RPA calculation is implemented using the ppf-fRG with Katanin truncation but a restriction to the terms including a site-sum $\sum_j$ on the right-hand side of Eqns.~\eqref{eq:dGamma_s}, \eqref{eq:dGamma_d} and \eqref{eq:SelfEnergyFlow}. The RPA confirms that the implementation of the PF term can reproduce the analytic spin mean-field result $T_c=1.5$ and $\nu=0.5$ for Heisenberg spins, see Fig.~\ref{fig:OtherTruncations} (a), (c). The one-loop scheme without Katanin truncation yields $T_c=1.29$ as well as a critical exponent consistent with the mean-field result $\nu=0.5$, see Fig \ref{fig:OtherTruncations} (b), (d). In both the RPA and the one-loop truncation without Katanin, the spin projection exceeds unity. 

\begin{figure}[h]
    \centering
    \subfloat{
\includegraphics{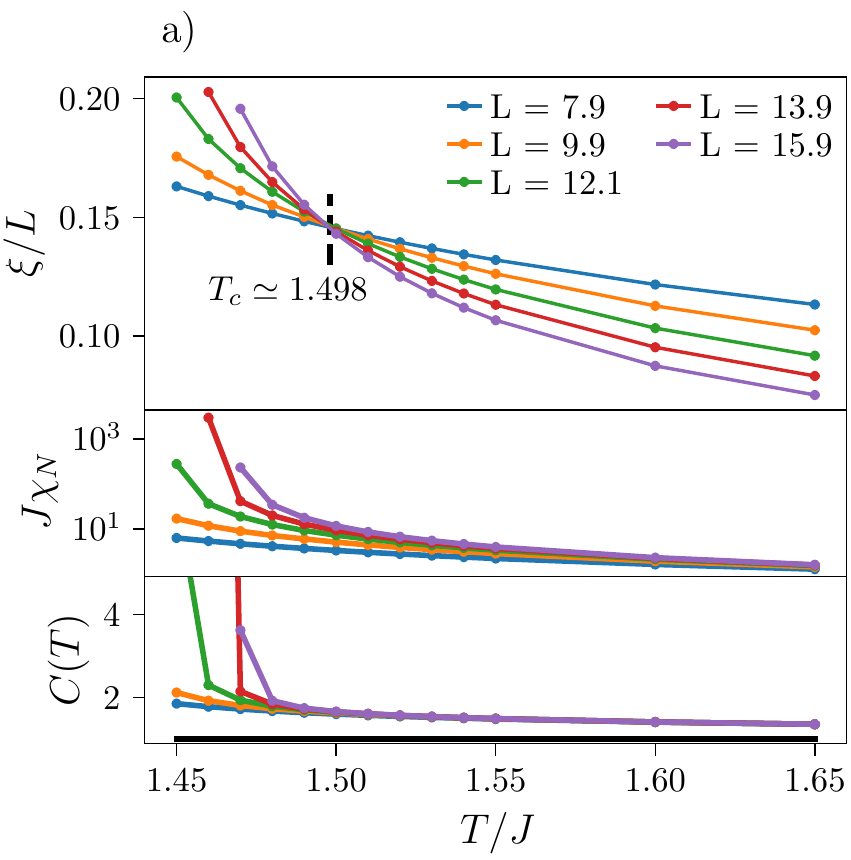}
    }\hfill
    \subfloat{
\includegraphics{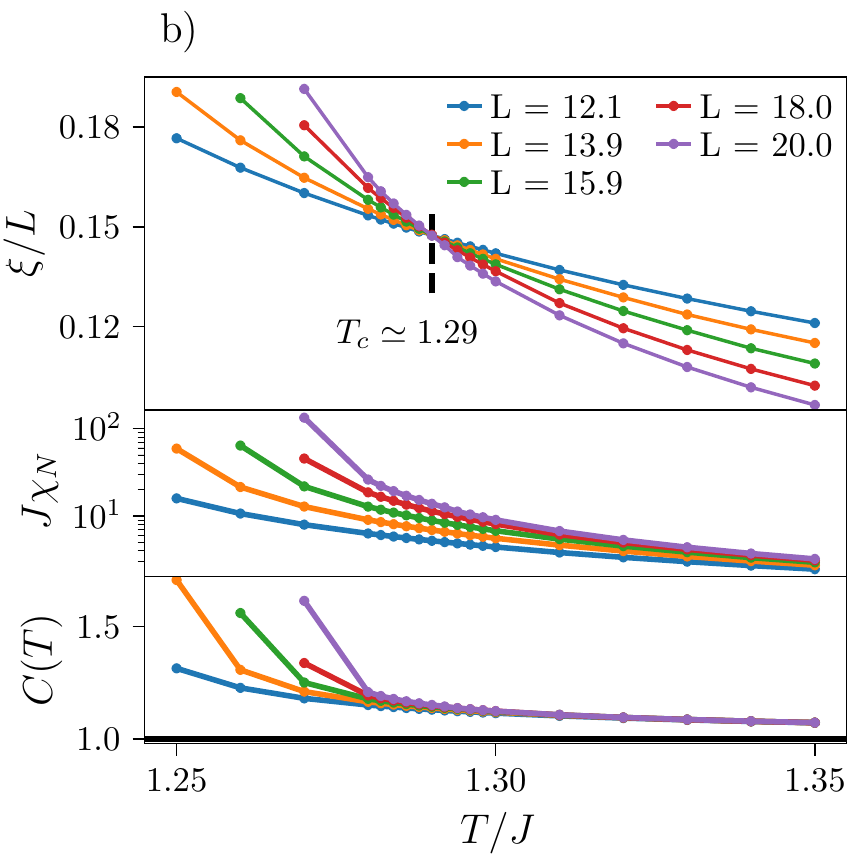}
}\\
    \subfloat{
\includegraphics{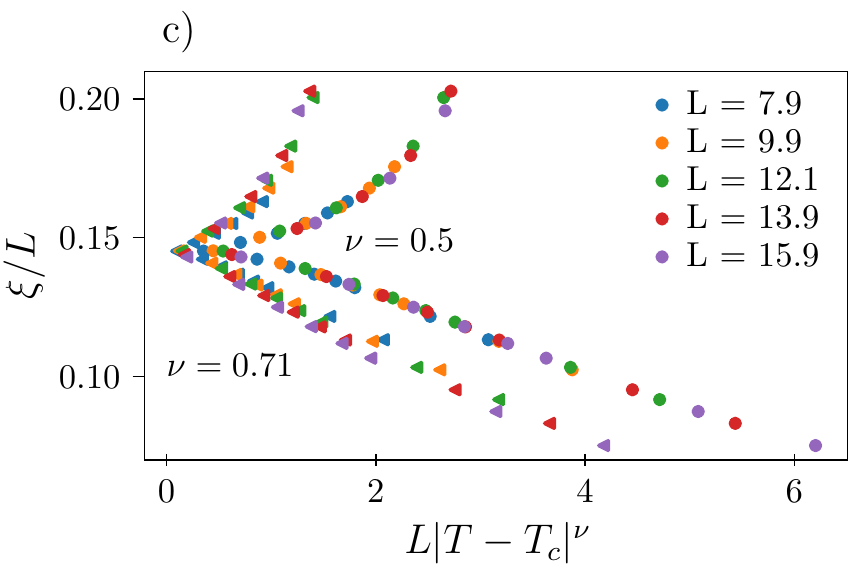}
    }\hfill
    \subfloat{
\includegraphics{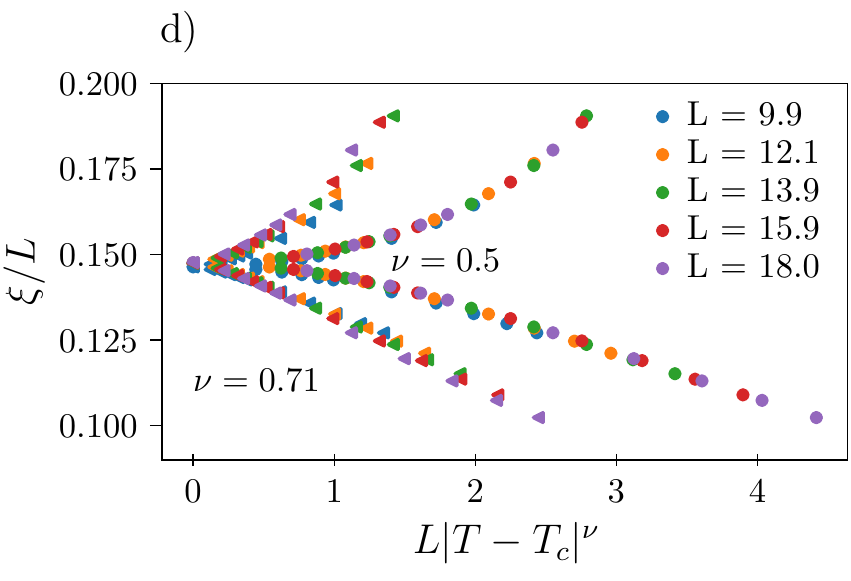}
    }
    \caption{Correlation length, N\'{e}el susceptibility, spin projection and scaling collapse from the calculations on the nearest neighbor AFM Heisenberg model on the cubic lattice. For panel (a) and (c) the flow equations were truncated via the RPA and integrated out numerically. We find $T_c\simeq 1.498$ and a critical exponent consistent with the mean-field value $\nu=0.5$. The spin projection deviates strongly from the exact result and is greater than 1.
    For panel (b) and (d) one-loop fRG without Katanin truncation was used. We find $T_c\simeq 1.29$ and a critical exponent also consistent with $\nu = 0.5$. The spin projection exceeds the exact result of $C(T)=1$.}
    \label{fig:OtherTruncations}
\end{figure}

We finally consider the ppf-PA formalism. As can be seen in the upper panel of Fig.~\ref{fig:ParquetCubicAFM}, we indeed find a smeared-out line crossing for $\xi/L$, with the three largest system sizes crossing at $T_c \simeq 0.855$. However, the corresponding N\'{e}el susceptibilities (see the middle panel in Fig.~\ref{fig:ParquetCubicAFM}) are substantially smaller than those obtained in ppf-fRG and a proper scaling collapse can be found neither for $\nu = 0.5$ nor $\nu = 0.71$. On the numerical side, we find that the rate of convergence drops considerably for $T / J \lesssim 1.0$ and the results, including the location of the intersection point of the $\xi$-scaling, become highly sensitive to numerical details such as the specific choice of mixing factors and solution algorithm. This could indicate that the ppf-PA fixed points in vicinity of the critical regime are strongly repulsive and hard to access in numerical calculations.
In conclusion, for the assessment of magnetic phase transition the ppf-PA seems less reliable and consistent than the one-loop ppf-fRG scheme.

\clearpage
\begin{figure}[t]
    \centering
    \includegraphics{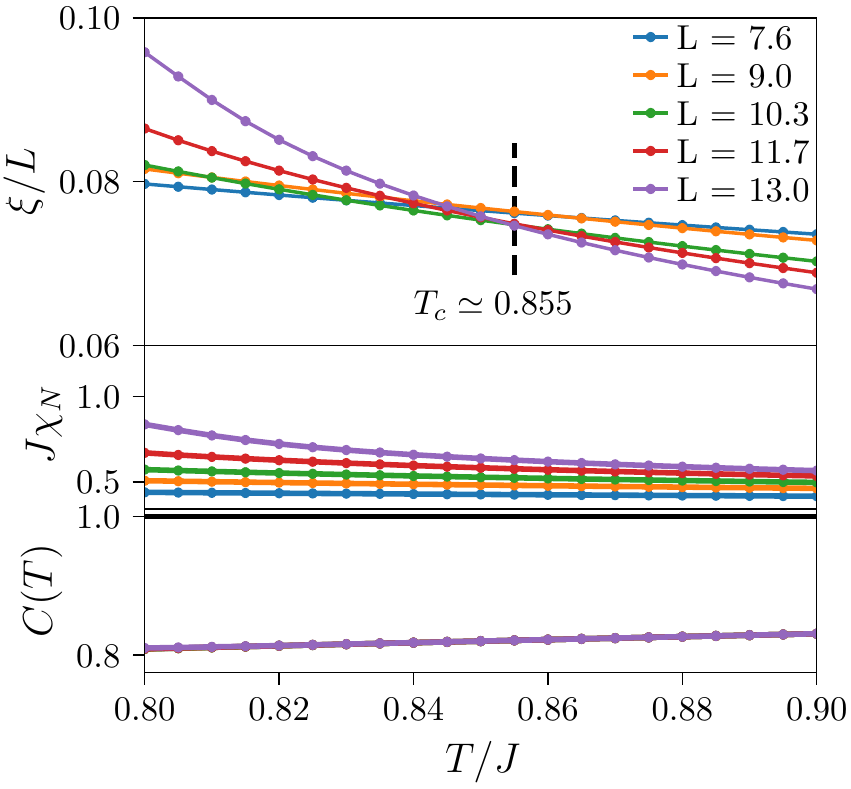}
    \caption{ppf-PA: Correlation length $\xi$, N\'{e}el susceptibility $\chi_N$ and spin projection $C$ for the nearest neighbor AFM Heisenberg model on the cubic lattice. We find a smeared-out line crossing at $T_c^\mathrm{ppf-PA} \simeq 0.855$. A corresponding scaling collapse consistent with $\nu=0.71$ or $\nu=0.5$ could not be found. For $C(T)$ all five lines lie on top of each other.}
    \label{fig:ParquetCubicAFM}
\end{figure}

\end{widetext}

\bibliographystyle{apsrev4-1}
\bibliography{Library}

\end{document}